# The Tully-Fisher relation : Correspondence between the Inverse and Direct approaches

S. Rauzy[1] and R. Triay[1*]

Université de Provence and Centre de Physique Théorique - C.N.R.S., Luminy Case 907, F-13288 Marseille Cedex 9, France.

**Abstract.** In a previous paper, we have demonstrated the importance to define a statistical model describing the observed linear correlation between the absolute magnitude $M$ and the log line width distance indicator $p$ of galaxies (the Tully-Fisher relation, Tully&Fisher (1977)). As long as the same statistical model is used during the calibration step of the relation and the step of the determination of the distances of galaxies, standard statistical methods such as the maximum likelihood technic permits us to derive bias free estimators of the distances of galaxies. However in practice, it is convenient to use a different statistical model for calibrating the Tully-Fisher relation (because of its robustness, the Inverse Tully-Fisher relation is prefered during this step) and for determining the distances of galaxies (the Direct Tully-Fisher relation is more accurate and robust in this case). Herein, we establish a correspondence between the Inverse and the Direct Tully-Fisher approaches. Assuming a gaussian luminosity function, we prove that the ITF and DTF models are in fact mathematically equivalent (i.e. they describe the same physical data distribution in the TF diagram). It thus turns out that as long as the calibration parameters are obtained for a given model, we can deduce the corresponding parameters of the other model. We present these formulae of correspondence and discuss their validitity for non-gaussian luminosity functions.

## 1. Introduction

Recently, the number of theoretical works on the Tully-Fisher (TF) like relation used to infer redshift independent estimators of distance of galaxies has increased rapidly (Hendry et al. (1994), Willick (1994), Sandage (1994),

Send offprint requests to: S. Rauzy
* the European Cosmological Network

Teerikorpi (1993), Bicknell (1992), Landy et al. (1992), Fouqué et al. (1990), Teerikorpi (1990), Hendry et al. (1990)). In a previous paper (Triay et al. (1994), hereafter TLR) we have demonstrated the importance to define a statistical model describing this observed linear correlation between the absolute magnitude $M$ and the log line width distance indicator $p$ of galaxies. A random variable $\zeta = a\,p + b - M$ of zero mean was introduced to mimic the intrinsic scatter $\sigma_\zeta$ of the TF relation. In order to fully specify the statistical model, a second random variable $\xi$ of mean $\xi_0$ and dispersion $\sigma_\xi$, statistically independent of $\zeta$, has to be chosen. The Inverse TF relation (ITF) or the Direct TF relation (DTF) appears indeed as a peculiar choice of this variable $\xi$. As long as the selection effects in observation are taken into account, we have shown that the introduction of such a statistical model permits to derive unbiased statistics for distance of galaxies as well as for the Hubble's constant by using standard statistical methods like the maximum likelihood technic (TLR and Triay et al. (1995), hereafter TRL). In this present paper, we focus on the infuence of the statistical model chosen to mimic the genuine TF relation in the $M$-$p$ plane. The problems of biases related to selection effects in observation, measurement errors or the spatial distribution of sources are investigated in TLR and TRL.

In section 2, we generalize some results obtained in TLR by introducing a class of statistical models indexed by an angle parameter $\alpha$. This class of $\alpha$-models forms a continuous set of models including the ITF and DTF approaches as boundary cases. We derive the maximum likelihood statistics for the 5 model-dependent parameters $a^\alpha$, $b^\alpha$, $\sigma_\zeta^\alpha$, $\xi_0^\alpha$ and $\sigma_\xi^\alpha$ characterising an $\alpha$-model and we illustrate their variations with respect to the angle parameter $\alpha$. Assuming standard working hypothesis (a gaussian luminosity function), we prove in section 3 that all these $\alpha$-models are indeed mathematically equivalent : i.e. they describe the same physical data distribution in the $M$-$p$ plane. It thus turns out that as long as the 5 parameters $a^\alpha$, $b^\alpha$, $\sigma_\zeta^\alpha$, $\xi_0^\alpha$ and $\sigma_\xi^\alpha$ are known for a given $\alpha$-model (say the ITF model for example), we can deduce the corre-for the DTF model). These formulae of correspondence are derived in section 4. This property permits indeed to use a different statistical model for calibrating the TF relation and for determining the distance of galaxies or the Hubble's constant. In practice, the best suitable model will thus be chosen with regard to the selection effects affecting the samples during these 2 steps. We analyse the general case (i.e : non-gaussian luminosity functions) in section 5. We show that the TF relation remains fully characterized by the knowledge of the one and second order moments in $M$ and $p$, but that it is no longer possible to describe the data with the ITF and DTF models simultaneously. This implies that it doesn't exist formulae of correspondence for the general case. However in practice, the ITF and DTF descriptions are found to be sufficiently accurate approximations as long as the correlation coefficient of the TF relation is close to 1, which is the case for the real data. Standard notations and useful formulae used throughout the text are given in appendix A.

## 2. The set of the $\alpha$-models

Herein, we specify the theoretical probability density (*pd*) describing the distribution of variables involved in the TF relation. These variables are related to intrinsic quantities of sources (galaxies), which are :

– the absolute magnitude $M$,
– the log line width distance estimator $p = \log W$.

Regardless of selection effects in observation, measurement errors and of the distribution of sources in space, the theoretical *pd* describing the distribution of these variables in the $M$-$p$ plane can be written :

$$dP_{\rm th} = F(M,p)\,dM\,dp \qquad (1)$$

The observed linear correlation between $M$ and $p$ (the TF relation) constrains the probability density function (*pdf*) $F(M,p)$ to adopt a specific form. In fact, it exists a straight line $\Delta_{TF}$ of equation $\widetilde{M}(p) = a\,p + b$ such that the data in the $M$-$p$ plane are distributed about this line. The slope $a$ and the zero point $b$ of this line are unknown quantities which will be estimated during a preliminar calibration step. In TLR we have shown that it is convenient to express this intrinsic scatter about the line $\Delta_{TF}$ by introducing a random variable $\zeta$ of <u>zero mean</u> and of dispersion $\sigma_\zeta$ defined as follows :

$$\zeta = \widetilde{M}(p) - M = a\,p + b - M \qquad (2)$$

A second random variable $\xi$ statistically independent of $\zeta$ is required in order to fully specify the statistical model (i.e. the *pdf* $F(M,p)$) characterizing the data distribution in the $M$-$p$ plane [1].

---

[1] In the absence of a better physical understanding of the TF relation, the parameters $a$ and $b$ have to be determinated

introducing a set of models characterized by the choice of this second variable $\xi$. We define a family of model dependent variables $\xi^\alpha$ indexed by an angle parameter $\alpha$ varying continuously from 0 to $\pi/2$ :

$$\xi^\alpha = \cos\alpha\,M + \sin\alpha\,a^\alpha p \qquad (3)$$

where we rewrite the Eq. (2) as follows ($\zeta$ is model dependent, see footnote 1) :

$$\zeta^\alpha = \widetilde{M}^\alpha(p) - M = a^\alpha p + b^\alpha - M \qquad (4)$$

The random variable $\xi^\alpha$ is a linear combination of $M$ and $p$ and is statistically independent of the random variable $\zeta^\alpha$ characterizing the TF relation :

$$\text{Cov}(\xi^\alpha, \zeta^\alpha) = 0 \qquad (5)$$

see *Def.* 4 in the appendix A. We have thus introduced a set of statistical $\alpha$-models describing the TF diagram by the following *pd* :

$$dP_{\rm th} \approx dP_{\rm th}^\alpha = f_{\xi^\alpha}(\xi^\alpha)d\xi^\alpha\ g(\zeta^\alpha;0,\sigma_\zeta^\alpha)d\zeta^\alpha \qquad (6)$$

In order to entirely characterize an $\alpha$-model, we need to specify the *pdf* $g(\zeta^\alpha;0,\sigma_\zeta^\alpha)$ and $f_{\xi^\alpha}(\xi^\alpha)$. The distribution of the random variable $\zeta^\alpha$ is chosen gaussian and we limit ourselves in this section to the case of a gaussian distribution for $\xi^\alpha$ (the general case is discussed section 5). Our <u>working hypothesis</u> are then a gaussian *pdf* of zero mean and of dispersion $\sigma_\zeta^\alpha$ for the random variable $\zeta^\alpha$ characterizing the intrinsic scatter about the straight line $\Delta_{TF}^\alpha$ and a gaussian *pdf* of mean $\xi_0^\alpha$ and of dispersion $\sigma_\xi^\alpha$ for the second random variable $\xi^\alpha$. Note that our $\alpha$-models are based on the independence of $\zeta^\alpha$ and $\xi^\alpha$, which means in particular that $\sigma_\zeta^\alpha$ doesn't depend on $\xi^\alpha$, and $\xi_0^\alpha$ and $\sigma_\xi^\alpha$ on $\zeta^\alpha$ :

$$g(\zeta^\alpha;0,\sigma_\zeta^\alpha) = g_G(\zeta^\alpha;0,\sigma_\zeta^\alpha) \qquad (7)$$
$$f_{\xi^\alpha}(\xi^\alpha) = g_G(\xi^\alpha;\xi_0^\alpha,\sigma_\xi^\alpha) \qquad (8)$$

see *Def.* 1 in the appendix A. Finally, the *pd* describing an $\alpha$-model reads as follows :

$$dP_{\rm th}^\alpha = g_G(\xi^\alpha;\xi_0^\alpha,\sigma_\xi^\alpha)d\xi^\alpha\ g_G(\zeta^\alpha;0,\sigma_\zeta^\alpha)d\zeta^\alpha \qquad (9)$$

Note that the set of the $\alpha$-models describes the Direct TF relation ($p$ and $\zeta$ are statistically independent) and the Inverse TF relation ($M$ and $\zeta$ are statistcally independent) when the angle parameter $\alpha$ is equal to its boundary values :

$$\text{ITF}:\begin{cases}\alpha = 0\\ dP_{\rm th}^{\rm I} = g_G(M;M_0,\sigma_M)dM\ g_G(\zeta^{\rm I};0,\sigma_\zeta^{\rm I})d\zeta^{\rm I}\end{cases} \qquad (10)$$

---

using a statistical process (the calibration step). Thus, these parameters $a$ and $b$ and so the random variables $\zeta$ and $\xi$ depend on the statistical model used to describe the data distribution in the $M$-$p$ plane : they are model dependent.

$$\text{DTF}: \begin{cases} dP_{\text{th}}^{\text{D}} = g_G(p; p_0, \sigma_p) dp \, g_G(\zeta^{\text{D}}; 0, \sigma_\zeta^{\text{D}}) d\zeta^{\text{D}} \end{cases} \quad (11)$$

The next step of the analysis is to derive the 5 model dependent parameters $a^\alpha$, $b^\alpha$, $\sigma_\zeta^\alpha$, $\xi_0^\alpha$ and $\sigma_\xi^\alpha$ characterising an $\alpha$-model from a calibration sample (containing $N$ galaxies with observed $M$ and $p$, $\{M_k, p_k\}_{k=1,N}$). Herein, we use the maximum likelihood technic (see *Def.* 3 in the appendix A) to derive these statistics which are presented in the appendix B. We illustrate in Fig. 1 the variation of these 5 model dependent parameters for $\alpha$ varying continuously from 0 to $\pi/2$. The simulated calibration sample is chosen large enough in order that averaged quantities are close to their expected values.

In particular the general statistics of appendix B for the ITF model ($\alpha = 0$) reads (with the correlation coefficient $\rho(p, M)$ defined in *Def.* 4 of the appendix A) :

$$a^{\text{I}} = \frac{\Sigma(M)^2}{\text{Cov}(p, M)} \quad (12)$$

$$b^{\text{I}} = \langle M \rangle - \frac{\Sigma(M)^2}{\text{Cov}(p, M)} \langle p \rangle \quad (13)$$

$$\sigma_\zeta^{\text{I}\,2} = \Sigma(M)^2 \left( \frac{1}{\rho(p, M)^2} - 1 \right) \quad (14)$$

$$\xi_0^{\text{I}} = \langle M \rangle \quad (15)$$

$$\sigma_\xi^{\text{I}\,2} = \Sigma(M)^2 \quad (16)$$

and for the DTF model ($\alpha = \pi/2$) :

$$a^{\text{D}} = \frac{\text{Cov}(p, M)}{\Sigma(p)^2} \quad (17)$$

$$b^{\text{D}} = \langle M \rangle - \frac{\text{Cov}(p, M)}{\Sigma(p)^2} \langle p \rangle \quad (18)$$

$$\sigma_\zeta^{\text{D}\,2} = \Sigma(M)^2 \left( 1 - \rho(p, M)^2 \right) \quad (19)$$

$$\xi_0^{\text{D}} = a^{\text{D}} \langle p \rangle = \frac{\text{Cov}(p, M)}{\Sigma(p)^2} \langle p \rangle \quad (20)$$

$$\sigma_\xi^{\text{D}\,2} = a^{\text{D}\,2} \Sigma(p)^2 = \rho(p, M)^2 \Sigma(M)^2 \quad (21)$$

### 3. Equivalence of the $\alpha$-models

In substituting the general statistics of the model dependent parameters $a^\alpha$, $b^\alpha$, $\sigma_\zeta^\alpha$, $\xi_0^\alpha$ and $\sigma_\xi^\alpha$ in the *pd* of Eq. (9), we find that, for every $\alpha$ belonging to $[0, \pi/2]$ (see appendix C for detailed calculations) :

$$\forall \alpha \in [0, \pi/2] \;:\; dP_{\text{th}}^\alpha =$$

$$g_G\left(M; \langle M \rangle + \frac{\text{Cov}(p,M)}{\Sigma(p)^2}(p - \langle p \rangle), \Sigma(M)\sqrt{1-\rho^2}\right) \quad (22)$$

$$\times g_G(p; \langle p \rangle, \Sigma(p)) \, dM \, dp$$

It thus means that all the $\alpha$-models are indeed mathematically equivalent. They describe the <u>same</u> physical distribution of data in the $M$-$p$ plane. Note that, by using rewrite Eq. (22) as follows :

$$\forall \alpha \in [0, \pi/2] \;:\; dP_{\text{th}}^\alpha =$$

$$g_G\left(p; \langle p \rangle + \frac{\text{Cov}(p,M)}{\Sigma(M)^2}(M - \langle M \rangle), \Sigma(p)\sqrt{1-\rho^2}\right) \quad (23)$$

$$\times g_G(M; \langle M \rangle, \Sigma(M)) \, dM \, dp$$

The symmetric expression in $M$ and $p$ of Eq. (22, 23) reads finally as a binormal *pdf* in $M$ and $p$, entirely characterized by its 5 moments of first and second order $\langle p \rangle$, $\langle M \rangle$, $\Sigma(p)$, $\Sigma(M)$ and $\text{Cov}(p, M)$ :

$$\forall \alpha \in [0, \pi/2] \;:\; dP_{\text{th}}^\alpha =$$

$$\frac{1}{2\pi \Sigma(M)\Sigma(p)\sqrt{1-\rho(p,M)^2}} \exp\left\{ -\frac{1}{2(1-\rho(p,M)^2)} \right.$$

$$\left. \times \left( \frac{(p-\langle p \rangle)^2}{\Sigma(p)^2} - 2\frac{\text{Cov}(p,M)(p-\langle p \rangle)(M-\langle M \rangle)}{\Sigma(M)^2\Sigma(p)^2} + \frac{(M-\langle M \rangle)^2}{\Sigma(M)^2} \right) \right\} \quad (24)$$

$$\times dM \, dp$$

We now understand that our working hypothesis (2 gaussian *pdf* for $\xi^\alpha$ and $\zeta^\alpha$) imply that the knowledge of the 5 parameters $a^\alpha$, $b^\alpha$, $\sigma_\zeta^\alpha$, $\xi_0^\alpha$ and $\sigma_\xi^\alpha$ for a given $\alpha$-model is sufficient to entirely describe the data distribution of the TF diagram [2].

The statistical problem underlying the Tully-Fisher relation is sometimes expressed in the literature in terms of conditional probability (see for example Hendry et al. (1994) or Lynden-Bell et al. (1988)). The Direct Tully-Fisher approach consists in defining for a given value of the parameter $p$, a mean absolute magnitude $\widetilde{M}^{\text{D}}(p)$ depending linearly on $p$ :

$$\widetilde{M}^{\text{D}}(p) = a^{\text{D}} p + b^{\text{D}} \quad (25)$$

where $a^{\text{D}}$ and $b^{\text{D}}$ (the slope and the zero point of the Direct Tully-Fisher line) correspond in our formalism to $a^\alpha$ and $b^\alpha$ with $\alpha = \pi/2$ (see Eq. (17, 18)). The conditional *pd* of $M$ given $p$ takes then the following form :

$$dP_{\text{th}}(M|p) = P_{\text{th}}(M|p) \, dM$$
$$= g_G(M; \widetilde{M}^{\text{D}}(p), \sigma_\zeta^{\text{D}}) \, dM \quad (26)$$

The distribution of the $M$'s is centered on $\widetilde{M}^{\text{D}}(p)$ and of dispersion $\sigma_\zeta^{\text{D}}$. Finally, the *pd* in the $M$-$p$ plane reads :

$$dP_{\text{th}} = g_G(p; \langle p \rangle, \Sigma(p)) P_{\text{th}}(M|p) \, dM \, dp \quad (27)$$

which is Eq. (22) when substituting the calibration parameters $a^{\text{D}}$, $b^{\text{D}}$ and $\sigma_\zeta^{\text{D}}$ by their estimates given in Eq.

---

[2] Weaker hypothesis on the 2 *pdf* oblige us to take into account the higher order moments of the bivariate distribution in $M$ and $p$. Thus, the $\alpha$-models are no longer strictly equivalent. This case is developed in section 5.

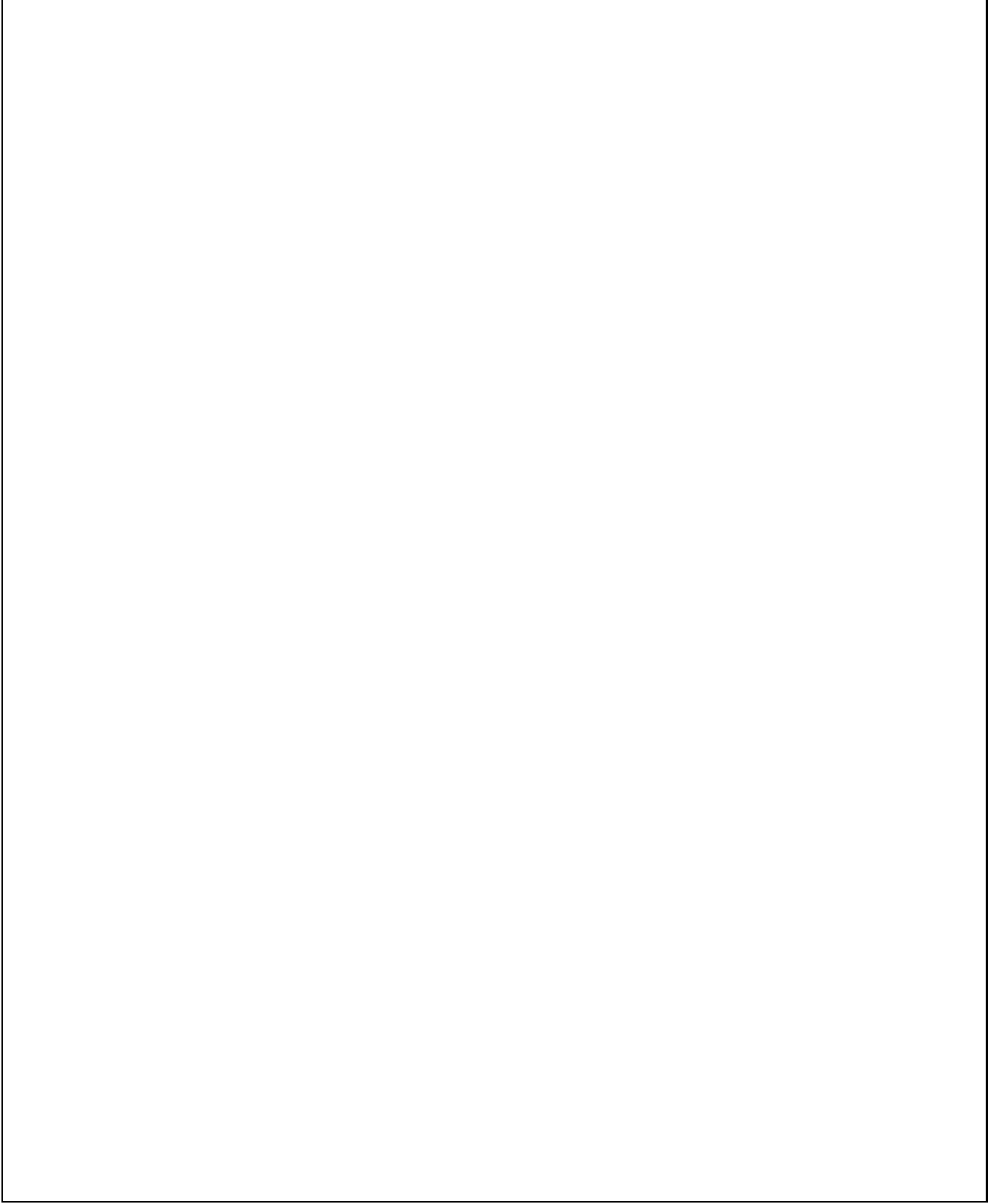

**Fig. 1.** Variations of the 5 model dependent parameters $a^\alpha$, $b^\alpha$, $\sigma_\zeta^\alpha$, $\xi_0^\alpha$ and $\sigma_\xi^\alpha$ with respect to the angle parameter $\alpha$.

is based on the introduction of a mean parameter $\tilde{p}^{\mathrm{I}}(M)$ for a given value of $M$:

$$\tilde{p}^{\mathrm{I}}(M) = \frac{1}{a^{\mathrm{I}}}M - \frac{b^{\mathrm{I}}}{a^{\mathrm{I}}} \tag{28}$$

with $a^{\mathrm{I}}$ and $b^{\mathrm{I}}$ the splope and the zero point of the Inverse Tully-Fisher line (see Eq. (12, 13)). The conditional pd of $p$ given $M$ is expressed as follows:

$$\begin{aligned} dP_{\mathrm{th}}(p|M) &= P_{\mathrm{th}}(p|M)\, dp \\ &= g_G(p; \tilde{p}^{\mathrm{I}}(M), \sigma_\zeta^{\mathrm{I}}/a^{\mathrm{I}})\, dp \end{aligned} \tag{29}$$

And the pd describing the data in the $M$-$p$ plane is derived from Eq. (23) by using Eq. (12, 13, 14).

$$dP_{\mathrm{th}} = g_G(M; \langle M \rangle, \Sigma(M)) P_{\mathrm{th}}(p|M)\, dM\, dp \tag{30}$$

## 4. Correspondence between the $\alpha$-models

We have shown in section 3 that, with our standard working hypothesis, all the $\alpha$-models are indeed mathematically equivalent. It thus turns out that if the 5 calibration parameters which describe the data distribution in the $M$-$p$ plane are known for a given $\alpha$-model, we can deduce the corresponding 5 parameters for every $\alpha$-models. We derive in appendix D these general formulae of correspondence. Herein, we present these formulae when the angles $\alpha$ and $\beta$ are equal to their boundary values. If the calibration parameters of the Inverse Tully-Fisher relation are known and we want to infer the calibration parameters of the Direct Tully-Fisher relation ($\alpha = 0$ and $\beta = \pi/2$):

$$a^{\mathrm{D}} = a^{\mathrm{I}} \frac{\sigma_\xi^{\mathrm{I}\,2}}{\sigma_\xi^{\mathrm{I}\,2} + \sigma_\zeta^{\mathrm{I}\,2}} = \rho^{\mathrm{I}\,2} a^{\mathrm{I}} \tag{31}$$

$$b^{\mathrm{D}} = \left(1 - \rho^{\mathrm{I}\,2}\right) \xi_0^{\mathrm{I}} + \rho^{\mathrm{I}\,2} b^{\mathrm{I}} \tag{32}$$

$$\xi_0^{\mathrm{D}} = \rho^{\mathrm{I}\,2} \left(\xi_0^{\mathrm{I}} - b^{\mathrm{I}}\right) \tag{33}$$

$$\sigma_\zeta^{\mathrm{D}\,2} = \left(1 - \rho^{\mathrm{I}\,2}\right)^2 \sigma_\xi^{\mathrm{I}\,2} + \rho^{\mathrm{I}\,4} \sigma_\zeta^{\mathrm{I}\,2} = \rho^{\mathrm{I}\,2} \sigma_\zeta^{\mathrm{I}\,2} \tag{34}$$

$$\sigma_\xi^{\mathrm{D}\,2} = \rho^{\mathrm{I}\,4} \left(\sigma_\xi^{\mathrm{I}\,2} + \sigma_\zeta^{\mathrm{I}\,2}\right) = \rho^{\mathrm{I}\,2} \sigma_\xi^{\mathrm{I}\,2} \tag{35}$$

Or conversely if the calibration parameters of the Direct Tully-Fisher relation are known and we want to deduce the calibration parameters of the Inverse Tully-Fisher relation ($\alpha = \pi/2$ and $\beta = 0$):

$$a^{\mathrm{I}} = a^{\mathrm{D}} \frac{\sigma_\xi^{\mathrm{D}\,2} + \sigma_\zeta^{\mathrm{D}\,2}}{\sigma_\xi^{\mathrm{D}\,2}} = \frac{1}{\rho^{\mathrm{D}\,2}} a^{\mathrm{D}} \tag{36}$$

$$b^{\mathrm{I}} = \left(1 - \frac{1}{\rho^{\mathrm{D}\,2}}\right) \xi_0^{\mathrm{D}} + b^{\mathrm{D}} \tag{37}$$

$$\xi_0^{\mathrm{I}} = \xi_0^{\mathrm{D}} + b^{\mathrm{D}} \tag{38}$$

$$\sigma_\zeta^{\mathrm{I}\,2} = \left(1 - \frac{1}{\rho^{\mathrm{D}\,2}}\right)^2 \sigma_\xi^{\mathrm{D}\,2} + \frac{1}{\rho^{\mathrm{D}\,4}} \sigma_\zeta^{\mathrm{D}\,2} = \frac{1}{\rho^{\mathrm{D}\,2}} \sigma_\zeta^{\mathrm{D}\,2} \tag{39}$$

$$\sigma_\xi^{\mathrm{I}\,2} = \sigma_\xi^{\mathrm{D}\,2} + \sigma_\zeta^{\mathrm{D}\,2} = \frac{1}{\rho^{\mathrm{D}\,2}} \sigma_\xi^{\mathrm{D}\,2} \tag{40}$$

## 5. Generalization and discussion

Assuming that the $M$-$p$ distribution is a binormal pdf, we have outlined in section 3 that the ITF and DTF approaches are indeed equivalent and that the Tully-Fisher relation in this special case is entirely characterized by the five moments (of first and second order) $\langle p \rangle$, $\langle M \rangle$, $\Sigma(p)$, $\Sigma(M)$ and $\mathrm{Cov}(p, M)$. However such a property vanishes if the luminosity function is not gaussian. We show in the appendix E that the ITF and DTF approaches are no longer strictly equivalent for general luminosity functions. In order to preserve this equivalence, the luminosity function $f_M(M)$ of the analysed galaxies has to verify the following equation:

$$f_M(M) \approx f_{rec}(M) = \int f_M(x) W_M(x)\, dx \tag{41}$$

with $\rho = \rho(M,p)$ the correlation coefficient of the data distribution and $W_M(x)$ given by (see appendix E):

$$\begin{aligned} W_M(x) &= \\ &\frac{1}{\rho^2} g_G\left(x; M + \frac{1-\rho^2}{\rho^2}(M - \langle M \rangle), \frac{\Sigma(M)}{\rho^2}\sqrt{1-\rho^2}\right) \end{aligned} \tag{42}$$

This condition is achieved when the luminosity function is gaussian but is ruled out in general. The errors introduced by describing indifferently the data distribution with the ITF or DTF models (or with some $\alpha$-model) can be quantified by using a Kolmogorov-Smirnov test as test of goodness of fit. The amplitude of these errors on the distance estimates of galaxies depends on the characteristics of the galaxies population (luminosity function, correlation coefficient of the TF relation, etc ...). Hence, we do not estimate numerically the amplitude of the errors and only speculate qualitatively on their general behaviour. We use the discrepancy between $f_M(M)$ and $f_{rec}(M)$ as a measure of the errors introduced by describing silmutaneously the data distribution with the ITF and DTF models. Eq (42,41) show that this discrepancy is negligible as long as the correlation coefficient of the TF relation is close to 1 (which is the case for the real data). We illustrate this feature (Fig. 2) by plotting $f_M(M)$ and $f_{rec}(M)$ for a Schechter luminosity function. We see that the condition of equation (41) is poorly verified for small correlation coefficient ($\rho = -0.85$). In this case, it becomes dangerous to use a priori a model to describe the Tully-Fisher relation. A bad choice of this model can create spurious phenomena such as the variation with $M$ or $p$ of the slope or the dispersion of the Tully-Fisher relation. Similar effects may also appear when calibrating the TF relation in distant clusters since the apparent magnitude cut-off favours the selection of high absolute magnitude galaxies for which the discrepancy between $f_M(M)$ and $f_{rec}(M)$ is large. This discrepancy however decreases when adopting for the correlation coefficient a more realistic value $\rho = -0.95$ (except for the high magnitudes, which is not a prejudice since the Schechter high magnitude cut-off is not physical). Because the $f_{rec}(M)$ curve for $\rho = -0.95$ match well the Schechter function, it seems that the ITF and DTF models (and so the set of the $\alpha$-models) are close to be equivalent for the real observed data. It follows that the formulae of correspondence derived in section 4 appears as fair approximations even for non-gaussian luminosity functions[3].

**Fig. 2.** Comparison between $f_M(M)$ and $f_{rec}(M)$ for a Schechter luminosity function.

## 6. Conclusion

In order to mimic the Tully-Fisher diagram, we have introduced a continuous set of statistical models characterized by the straight line $\Delta_{TF}^\alpha$ describing the observed linear correlation of $M$ and $p$. This set of $\alpha$-models includes the ITF and DTF relation as boundary cases. Assuming a gaussian luminosity function, we have shown that all these $\alpha$-models describe indeed the same physical data distribution in the $M$-$p$ plane. Thus, if the 5 calibration parameters $a^\alpha$, $b^\alpha$, $\sigma_\zeta^\alpha$, $\xi_0^\alpha$ and $\sigma_\xi^\alpha$ are known for a given $\alpha$-model, we can infer the calibration parameters of every $\alpha$-models by using formulae of correspondence. We have shown that this property remains valid for non-gaussian luminosity functions as long as the correlation coefficient of the TF diagram is close to 1 (which is the case for the real data). In practice such a property allows us the possibility to use a different statistical model during the calibration step of the TF relation and for determining the distances of galaxies. The best suitable statistical model will thus be chosen with regard to the selection effects in observation affecting the samples during each of these 2 steps.

For example, the ITF model seems to be more adequate for calibrating the TF relation because of its robustness (the estimates of $a^I$, $b^I$ and $\sigma_\zeta^I$ do not depend on the luminosity function (Hendry et al. (1990), TLR) but also because when calibrating the ITF relation in a cluster, the estimates of $a^I$ and $\sigma_\zeta^I$ don't depend on the distance of the cluster (Schechter (1980), Tully (1988), Lynden-Bell et al. (1988),Teerikorpi (1990), Hendry et al. (1990), Rauzy preferred to determine the distances of galaxies. It is more accurate (the intrinsic scatter of the DTF relation $\sigma_\zeta^D$ is indeed smaller than the ITF one $\sigma_\zeta^I$ (Tully (1988), TLR)), more robust (the DTF distance estimator doesn't depend on the luminosity function (TLR)) and more intuitive (an observed $p$ gives directly a value of $M$ : $\widetilde{M}(p) = a^D p + b^D$ (Bottinelli et al. (1986), Fouqué et (al. 1990))).

---

[3] Note however that if the shape of the luminosity function is of no importance during the calibration step, it can be decisive for the bias correction procedure involved in the determination of the distances of galaxies (see Bicknell (1992)).

## A. Notations and useful formulae

The mathematical formalism is similar to the one used in Bigot & Triay (1990). The following features are addressed throughout the text by using the symbol "*Def.*".

*Def.*1 The *probability density* (*pd*) of a random variable $x$ reads $dP(x) = f(x)dx$, where $f(x)$ represents the *pd function* (*pdf*), we have $\int dP(x) = 1$. Sometimes, it is useful to exhibit the *model parameters* involved in the statistical model, as the mean $x_0$ and the standard deviation $\sigma$, by writing $f(x; x_0, \sigma)$.
  (a) $g_G(x; x_0, \sigma) = (\sigma\sqrt{2\pi})^{-1} \exp - ((x - x_0)^2/(2\sigma^2))$ is a Gaussian *pdf*.
  (b) A normal *pdf* can be written $g_N(x) = g_G(x; 0, 1)$.

*Def.*2 Let $f$ be a *pdf*, and $\lambda$ be a scalar value, in most of calculations, we use the following properties :
  (a) $f(x + \lambda; x_0, \sigma) = f(x; x_0 - \lambda, \sigma)$;
  (b) $f(\lambda x; x_0, \sigma) = \lambda^{-1} f(x; \frac{x_0}{\lambda}, \frac{\sigma}{\lambda})$;
  (c) $g_G(x; x_1, \sigma_1) g_G(x; x_2, \sigma_2) = g_G(x; x_0, \sigma_0) g_G(x_1; x_2, \acute{\sigma})$, where $\acute{\sigma} = \sqrt{\sigma_1^2 + \sigma_2^2}$, $x_0$ and $\sigma_0$ are defined as follows $\sigma_0^{-2} = \sigma_1^{-2} + \sigma_2^{-2}$ and $x_0 \sigma_0^{-2} = x_1 \sigma_1^{-2} + x_2 \sigma_2^{-2}$.

*Def.*3 The *pd* of a sample data $\{\mathcal{G}_k\}_{k=1,N}$, which consists of $N$ independently selected objects $\mathcal{G}_k$, is given by $\prod_{k=1}^N dP(\mathcal{G}_k)$.
  (a) Its *pdf*, written in terms of *observables* (the measurable random variables), but regarded as a function of model parameters, provides us with the *likelihood function*.
  (b) (*The* ML *method*.) The model parameters statistics are obtained by maximizing the likelihood function, or (equivalently) the natural logarithm of the *efficient part* of it, in which the terms which do not contribute to the determination of parameters are removed, herein briefly denoted by *lf*.

*Def.*4 We use the following usual definitions :
  (a) $\langle x \rangle = \sum_{k=1}^N x_k / N$ is the average,
  (b) $\text{Cov}(x, y) = \sum_{k=1}^N (x_k - \langle x \rangle)(y_k - \langle y \rangle)/(N - 1)$ is the covariance,
  (c) $\Sigma(x) = \sqrt{\text{Cov}(x, x)}$ is the standard deviation,
  (d) $\rho(x, y) = \text{Cov}(x, y)/(\Sigma(x)\Sigma(y))$ is the correlation coefficient.

We have to express the *pd* of Eq. (9) in terms of the observables $M$ and $p$. In substituting Eq. (3,4) in Eq. (9), we obtain :

$$dP_{\text{th}}^\alpha = \quad g_G(c^\alpha M + s^\alpha a^\alpha p; \xi_0^\alpha, \sigma_\xi^\alpha)$$
$$\times g_G(a^\alpha p + b^\alpha - M; 0, \sigma_\zeta^\alpha) \qquad (B1)$$
$$\times |a^\alpha (c^\alpha + s^\alpha)| \, dM \, dp$$

where $c^\alpha = \cos\alpha$, $s^\alpha = \sin\alpha$ and $|a^\alpha(c^\alpha + s^\alpha)|$ is the Jacobian of the coordinate transformation. The natural logarithm of the likelihood function (*lf*) $\mathcal{L}^\alpha = \mathcal{L}^\alpha(a^\alpha, b^\alpha, \sigma_\zeta^\alpha, \xi_0^\alpha, \sigma_\xi^\alpha)$ as a function of the 5 parameters of the model reads :

$$\mathcal{L}^\alpha = \quad +\ln|a^\alpha| + \ln(c^\alpha + s^\alpha)$$
$$- \ln \sigma_\xi^\alpha - \frac{1}{N} \sum_{k=1}^N \frac{(c^\alpha M + s^\alpha a^\alpha p - \xi_0^\alpha)^2}{2{\sigma_\xi^\alpha}^2} \qquad (B2)$$
$$- \ln \sigma_\zeta^\alpha - \frac{1}{N} \sum_{k=1}^N \frac{(a^\alpha p + b^\alpha - M)^2}{2{\sigma_\zeta^\alpha}^2}$$

Maximizing $\mathcal{L}^\alpha$ with respect to these 5 parameters furnish the following set of equations (see *Def.* 3 and *Def.* 4 of the appendix A) :

$$\partial_{a^\alpha} \mathcal{L}^\alpha = 0 \Rightarrow \begin{cases} -\frac{1}{\sigma_\xi^{\alpha\,2}} \langle s^\alpha p (c^\alpha M + s^\alpha a^\alpha p - \xi_0^\alpha) \rangle \\ \\ -\frac{1}{\sigma_\zeta^{\alpha\,2}} \langle p(a^\alpha p + b^\alpha - M) \rangle + \frac{1}{a^\alpha} = 0 \end{cases} \quad (B3)$$

$$\partial_{b^\alpha} \mathcal{L}^\alpha = 0 \Rightarrow \langle (a^\alpha p + b^\alpha - M) \rangle = 0 \qquad (B4)$$

$$\partial_{\sigma_\zeta^{\alpha\,2}} \mathcal{L}^\alpha = 0 \Rightarrow \begin{cases} +\frac{1}{2\sigma_\zeta^{\alpha\,4}} \langle (a^\alpha p + b^\alpha - M)^2 \rangle \\ \\ -\frac{1}{2\sigma_\zeta^{\alpha\,2}} = 0 \end{cases} \qquad (B5)$$

$$\partial_{\xi_0^\alpha} \mathcal{L}^\alpha = 0 \Rightarrow \langle (c^\alpha M + s^\alpha a^\alpha p - \xi_0^\alpha) \rangle = 0 \qquad (B6)$$

$$\partial_{\sigma_\xi^{\alpha\,2}} \mathcal{L}^\alpha = 0 \Rightarrow \begin{cases} +\frac{1}{2\sigma_\xi^{\alpha\,4}} \langle (c^\alpha M + s^\alpha a^\alpha p - \xi_0^\alpha)^2 \rangle \\ \\ -\frac{1}{2\sigma_\xi^{\alpha\,2}} = 0 \end{cases} \qquad (B7)$$

Solving this set of equations provides the estimates of the 5 model dependent parameters $a^\alpha$, $b^\alpha$, $\sigma_\zeta^\alpha$, $\xi_0^\alpha$ and $\sigma_\xi^\alpha$ characterising an $\alpha$-model. Eq. (B4) furnishes the statistic of $b^\alpha$ :

$$b^\alpha = \langle M \rangle - a^\alpha \langle p \rangle \qquad (B8)$$

Substituting Eq. (B8) in Eq. (B5) gives :

$$\sigma_\zeta^{\alpha\,2} = \Sigma(M)^2 - 2a^\alpha \text{Cov}(p, M) + a^{\alpha\,2} \Sigma(p)^2 \qquad (B9)$$

Eq. (B6) gives the statistic of $\xi_0^\alpha$ :

$$\xi_0^\alpha = c^\alpha \langle M \rangle + s^\alpha a^\alpha \langle p \rangle \qquad (B10)$$

$$\sigma_\xi^{\alpha\,2} = c^{\alpha\,2}\Sigma(M)^2 + 2s^\alpha c^\alpha a^\alpha \mathrm{Cov}(p,M) \\ + s^{\alpha\,2}a^{\alpha\,2}\Sigma(p)^2 \tag{B11}$$

In order to evaluate the $a^\alpha$ statistic, we define $X = a^\alpha(p-\langle p\rangle)$ and $Y = (M-\langle M\rangle)$ (with these notations Eq. (B9) and Eq. (B11) rewrite $\sigma_\zeta^{\alpha\,2} = \langle(X-Y)^2\rangle$ and $\sigma_\xi^{\alpha\,2} = \langle(s^\alpha X - c^\alpha Y)^2\rangle$). In substituting Eq. (B8,B9,B10,B11) in Eq. (B3), we obtain (with $X$ and $Y$ defined above):

$$1 - \frac{\langle X(X-Y)\rangle}{\langle (X-Y)^2\rangle} - \frac{\langle s^\alpha X(s^\alpha X - c^\alpha Y)\rangle}{\langle (s^\alpha X - c^\alpha Y)^2\rangle} = 0 \tag{B12}$$

which reduces to:

$$\begin{cases} \langle s^\alpha X(s^\alpha X - c^\alpha Y)\rangle \langle (X-Y)^2\rangle \\ + \langle Y(X-Y)\rangle \langle (s^\alpha X - c^\alpha Y)^2\rangle = 0 \end{cases} \tag{B13}$$

and by developing and factorizing this equation:

$$\begin{cases} \left(s^\alpha \Sigma(X)^2 - c^\alpha \Sigma(Y)^2 + (c^\alpha - s^\alpha)\mathrm{Cov}(X,Y)\right) \\ \times \left(c^{\alpha\,2}\Sigma(Y)^2 + s^{\alpha\,2}\Sigma(X)^2\right) = 0 \end{cases} \tag{B14}$$

which rewrites:

$$\begin{cases} \left(a^{\alpha\,2}s^\alpha \Sigma(p)^2 - c^\alpha \Sigma(M)^2 + a^\alpha(c^\alpha - s^\alpha)\mathrm{Cov}(p,M)\right) \\ \times \left(c^{\alpha\,2}\Sigma(M)^2 + a^{\alpha\,2}s^{\alpha\,2}\Sigma(p)^2\right) = 0 \end{cases} \tag{B15}$$

which reduces to [4]:

$$a^{\alpha\,2}s^\alpha\Sigma(p)^2 + a^\alpha(c^\alpha-s^\alpha)\mathrm{Cov}(p,M) - c^\alpha\Sigma(M)^2 = 0 \tag{B16}$$

The $a^\alpha$ statistic is obtained straightforwardly if the coefficient of the second order term of Eq. (B16) vanishes:

$$\text{if } \alpha = 0 : a^\alpha = \frac{\Sigma(M)^2}{\mathrm{Cov}(p,M)} \tag{B17}$$

or else the $a^\alpha$ statistic is obtained by solving the second order equation of Eq. (B16):

if $\alpha \neq 0$ : $a^\alpha =$

$$\frac{1}{2s^\alpha\Sigma(p)^2}\left\{(s^\alpha - c^\alpha)\mathrm{Cov}(p,M) + \mathrm{sign}\left[\mathrm{Cov}(p,M)\right]\right. \tag{B18}$$

$$\left. \times \sqrt{(s^\alpha - c^\alpha)^2\mathrm{Cov}(p,M)^2 + 4c^\alpha s^\alpha \Sigma(M)^2\Sigma(p)^2}\right\}$$

We have thus derived the statistics providing an estimate of the 5 parameters $a^\alpha$, $b^\alpha$, $\sigma_\zeta^\alpha$, $\xi_0^\alpha$ and $\sigma_\xi^\alpha$ characterizing an $\alpha$-model.

---

[4] this equation can also be obtained by developing Eq. (5).

Herein we show, by substituting the general statistics of the model dependent parameters $a^\alpha$, $b^\alpha$, $\sigma_\zeta^\alpha$, $\xi_0^\alpha$ and $\sigma_\xi^\alpha$ in the pd of Eq. (9), that all the $\alpha$-models are indeed mathematically equivalent. For every $\alpha$ belonging to $[0,\pi/2]$ and by using the properties (a), (b) and (c) of the Def. 2 of the appendix A, we rewrite Eq. (9) as follows:

$$\begin{aligned} dP_{\mathrm{th}}^\alpha &= g_G(M; \tfrac{\xi_0^\alpha - s^\alpha a^\alpha p}{c^\alpha}, \tfrac{\sigma_\xi^\alpha}{c^\alpha}) \\ &\times g_G(M; a^\alpha p + b^\alpha, \sigma_\zeta^\alpha)\, \tfrac{|a^\alpha(c^\alpha+s^\alpha)|}{c^\alpha}\, dM\,dp \end{aligned} \tag{C1}$$

And by introducing the variables $x_0$, $\sigma_x$, $y_0$ and $\sigma_y$, we express the pd of Eq. (9) as the product of 2 gaussians depending on $M$ and $p$:

$$\begin{aligned} dP_{\mathrm{th}}^\alpha &= g_G\!\left(\tfrac{(c^\alpha+s^\alpha)}{c^\alpha}a^\alpha p; \tfrac{\xi_0^\alpha - s^\alpha a^\alpha p}{c^\alpha}, \sqrt{\sigma_\zeta^\alpha + \left(\tfrac{\sigma_\xi^\alpha}{c^\alpha}\right)^2}\right) \\ &\times g_G(M; x_0, \sigma_x)\, \tfrac{|a^\alpha(c^\alpha+s^\alpha)|}{c^\alpha}\, dM\,dp \end{aligned} \tag{C2}$$

$$dP_{\mathrm{th}}^\alpha = g_G(p; y_0, \sigma_y)g_G(M; x_0, \sigma_x)\, dM\,dp \tag{C3}$$

The variables $x_0$, $\sigma_x$, $y_0$ and $\sigma_y$ thus verify the following equalities:

$$\sigma_x^2 = \frac{\sigma_\zeta^{\alpha\,2}\sigma_\xi^{\alpha\,2}}{c^{\alpha\,2}\sigma_\zeta^{\alpha\,2} + \sigma_\xi^{\alpha\,2}} \tag{C4}$$

$$x_0 = \frac{c^\alpha \sigma_\zeta^{\alpha\,2}(\xi_0^\alpha - s^\alpha a^\alpha p) + \sigma_\xi^{\alpha\,2}(a^\alpha p + b^\alpha)}{c^{\alpha\,2}\sigma_\zeta^{\alpha\,2} + \sigma_\xi^{\alpha\,2}} \tag{C5}$$

$$\sigma_y^2 = \frac{1}{a^{\alpha\,2}(c^\alpha + s^\alpha)^2}\left(c^{\alpha\,2}\sigma_\zeta^{\alpha\,2} + \sigma_\xi^{\alpha\,2}\right) \tag{C6}$$

$$y_0 = \frac{\xi_0^\alpha - c^\alpha b^\alpha}{a^\alpha(c^\alpha + s^\alpha)} \tag{C7}$$

by replacing the statistics of $a^\alpha$, $b^\alpha$, $\sigma_\zeta^\alpha$, $\xi_0^\alpha$ and $\sigma_\xi^\alpha$ derived in Eq. (B8, B9, B10, B11, B17, B18), we find that for every $\alpha$ belonging to $[0,\pi/2]$:

$$\sigma_x^2 = \frac{\Sigma(M)^2\Sigma(p)^2 - \mathrm{Cov}(p,M)^2}{\Sigma(p)^2} \tag{C8}$$

$$= \Sigma(M)^2\left(1 - \rho(p,M)^2\right) \tag{C9}$$

$$x_0 = \langle M\rangle + \frac{\mathrm{Cov}(p,M)}{\Sigma(p)^2}(p - \langle p\rangle) \tag{C10}$$

$$\sigma_y^2 = \Sigma(p)^2 \tag{C11}$$

$$y_0 = \langle p\rangle \tag{C12}$$

And substituting Eq. (C8, C10, C11, C12) in Eq. (C3) gives Eq. (22) as required.

### D. General formulae of correspondence

The starting point is to assume that $a^\alpha$, $b^\alpha$, $\sigma_\zeta^\alpha$, $\xi_0^\alpha$ and $\sigma_\xi^\alpha$ are known for a given $\alpha$-model. By inverting the system

$$\Sigma(p)^2 = \frac{c^{\alpha\,2}\sigma_\zeta^{\alpha\,2} + \sigma_\xi^{\alpha\,2}}{a^{\alpha\,2}(c^\alpha + s^\alpha)^2} \tag{D1}$$

$$\mathrm{Cov}(p,M) = \frac{\sigma_\xi^{\alpha\,2} - c^\alpha s^\alpha \sigma_\zeta^{\alpha\,2}}{a^{\alpha\,2}(c^\alpha + s^\alpha)^2} \tag{D2}$$

$$\Sigma(M)^2 = \frac{s^{\alpha\,2}\sigma_\zeta^{\alpha\,2} + \sigma_\xi^{\alpha\,2}}{(c^\alpha + s^\alpha)^2} \tag{D3}$$

$$\langle M \rangle = \frac{\xi_0^\alpha + s^\alpha b^\alpha}{c^\alpha + s^\alpha} \tag{D4}$$

$$\langle p \rangle = \frac{\xi_0^\alpha - c^\alpha b^\alpha}{a^\alpha(c^\alpha + s^\alpha)} \tag{D5}$$

And so we can deduce the 5 parameters $a^\beta, b^\beta, \sigma_\zeta^\beta, \xi_0^\beta$ and $\sigma_\xi^\beta$ for every angle parameter $\beta$ belonging to $[0, \pi/2]$, by replacing the values of $\langle p\rangle, \langle M\rangle, \Sigma(p), \Sigma(M)$ and $\mathrm{Cov}(p,M)$ given in Eq. (D5, D4, D1, D3, D2), in Eq. (B8, B9, B10, B11, B17, B18) applied to a $\beta$-model. In the peculiar case $\beta = 0$, Eq. (B17) gives :

$$\text{if } \beta = 0 \;:\; a^\beta = a^\alpha \times \frac{s^{\alpha\,2}\sigma_\zeta^{\alpha\,2} + \sigma_\xi^{\alpha\,2}}{\sigma_\xi^{\alpha\,2} - c^\alpha s^\alpha \sigma_\zeta^{\alpha\,2}} \tag{D6}$$

And for the general case, we deduce $a^\beta$ from Eq. (B18). With $\mathrm{sgn} = \mathrm{sign}[\sigma_\xi^{\alpha\,2} - c^\alpha s^\alpha \sigma_\zeta^{\alpha\,2}]$, we obtain :

if $\beta \neq 0 \;:\; a^\beta = a^\alpha \times$

$$\frac{1}{2s^\beta(\sigma_\xi^{\alpha\,2} + c^{\alpha\,2}\sigma_\zeta^{\alpha\,2})} \times \Big\{ (s^\beta - c^\beta)(\sigma_\xi^{\alpha\,2} - c^\alpha s^\alpha \sigma_\zeta^{\alpha\,2})$$

$$+ \mathrm{sgn} \times \Big[(s^\beta - c^\beta)^2(\sigma_\xi^{\alpha\,2} - c^\alpha s^\alpha \sigma_\zeta^{\alpha\,2})^2 \tag{D7}$$

$$+ 4c^\beta s^\beta(s^{\alpha\,2}\sigma_\zeta^{\alpha\,2} + \sigma_\xi^{\alpha\,2})(c^{\alpha\,2}\sigma_\zeta^{\alpha\,2} + \sigma_\xi^{\alpha\,2})\Big]^{1/2}\Big\}$$

Finally, $b^\beta, \sigma_\zeta^\beta, \xi_0^\beta$ and $\sigma_\xi^\beta$ are derived using Eq. (B8, B9, B10, B11). We set $A^\alpha = \frac{1}{c^\alpha + s^\alpha}$ :

$$b^\beta = A^\alpha \left(\xi_0^\alpha(1 - \frac{a^\beta}{a^\alpha}) + b^\alpha(s^\alpha + c^\alpha\frac{a^\beta}{a^\alpha})\right) \tag{D8}$$

$$\xi_0^\beta = A^\alpha \left(\xi_0^\alpha(c^\beta + s^\beta\frac{a^\beta}{a^\alpha}) + b^\alpha(c^\beta s^\alpha - s^\beta c^\alpha\frac{a^\beta}{a^\alpha})\right) \tag{D9}$$

$$\sigma_\zeta^{\beta\,2} = A^{\alpha\,2}\left(\sigma_\xi^{\alpha\,2}(1 - \frac{a^\beta}{a^\alpha})^2 + \sigma_\zeta^{\alpha\,2}(s^\alpha + c^\alpha\frac{a^\beta}{a^\alpha})^2\right) \tag{D10}$$

$$\sigma_\xi^{\beta\,2} = A^{\alpha\,2}\left(\sigma_\xi^{\alpha\,2}(c^\beta + s^\beta\frac{a^\beta}{a^\alpha})^2 + \sigma_\zeta^{\alpha\,2}(c^\beta s^\alpha - s^\beta c^\alpha\frac{a^\beta}{a^\alpha})^2\right) \tag{D11}$$

### E. Non-Gaussian luminosity functions :

Let us assume that the ITF model describes the data distribution of some sample of galaxies characterized by a minosity function for example). The theoretical probability density $pd$ of a galaxy reads thus as follows :

$$dP_{\mathrm{th}} = f_M(M)dM \; g_G(\zeta^{\mathrm{I}}; 0, \sigma_\zeta^{\mathrm{I}})d\zeta^{\mathrm{I}} \tag{E1}$$

Is it now possible to describe this same data distribution by using the DTF model ?. If so, the theoretical $pd$ of Eq. (E1) can also be written :

$$dP_{\mathrm{th}} = g_p(p)dp \; g_G(\zeta^{\mathrm{D}}; 0, \sigma_\zeta^{\mathrm{D}})d\zeta^{\mathrm{D}} \tag{E2}$$

where $g_p(p)$ is the distribution function of the $p$'s and can be obtained by integrating over $M$ the $pdf$ of Eq. (E1). In the same way, the luminosity function $f_M(M)$ is equal to the integral over $p$ of the $pdf$ of Eq. (E2). It means that if we want to describe simultaneously the data distribution by Eq. (E1) and Eq. (E2), the luminosity function has to satisfy the following equation :

$$f_M(M) = f_{rec}(M) = \int f_M(x) W_M(x) dx \tag{E3}$$

with $W_M(x)$ given by :

$$W_M(x) = a^{\mathrm{I}} \int dp \quad g_G(x - a^{\mathrm{I}}p - b^{\mathrm{I}}; 0, \sigma_\zeta^{\mathrm{I}}) \\ \times g_G(M - a^{\mathrm{D}}p - b^{\mathrm{D}}; 0, \sigma_\zeta^{\mathrm{D}}) \tag{E4}$$

By using the properties (a), (b) and (c) of *Def. 2*, this equation rewrites :

$$W_M(x) = \frac{a^{\mathrm{I}}}{a^{\mathrm{D}}} \times \\ g_G\left(x; M + (\frac{a^{\mathrm{I}}}{a^{\mathrm{D}}} - 1)M + b^{\mathrm{I}} - \frac{a^{\mathrm{I}}}{a^{\mathrm{D}}}b^{\mathrm{D}}, \sqrt{\sigma_\zeta^{\mathrm{D}\,2} + (\frac{a^{\mathrm{I}}}{a^{\mathrm{D}}}\sigma_\zeta^{\mathrm{D}})^2}\right) \tag{E5}$$

Because the random variables $M$ and $\zeta^{\mathrm{I}}$ are independent in Eq. (E1), the statistics given in Eq. (12, 13, 14) of $a^{\mathrm{I}}, b^{\mathrm{I}}$ and $\sigma_\zeta^{\mathrm{I}}$ obtained by using the maximum likelihood technic in the appendix B remains valid. In the same way, $p$ and $\zeta^{\mathrm{D}}$ are independent in Eq. (E2), and so the statistics given in Eq. (17, 18, 19) still furnish us the estimates of $a^{\mathrm{D}}, b^{\mathrm{D}}$ and $\sigma_\zeta^{\mathrm{D}}$. It implies that the following equations holds :

$$\frac{a^{\mathrm{I}}}{a^{\mathrm{D}}} = \frac{\Sigma(M)^2 \Sigma(p)^2}{\mathrm{Cov}(p,M)^2} = \frac{1}{\rho(p,M)^2} \tag{E6}$$

$$b^{\mathrm{I}} - \frac{a^{\mathrm{I}}}{a^{\mathrm{D}}}b^{\mathrm{D}} = \left(1 - \frac{1}{\rho(M,p)^2}\right)\langle M \rangle \tag{E7}$$

$$\sigma_\zeta^{\mathrm{D}\,2} + \left(\frac{a^{\mathrm{I}}}{a^{\mathrm{D}}}\sigma_\zeta^{\mathrm{D}}\right)^2 = \Sigma(M)^2\left(\frac{1}{\rho(M,p)^4} - 1\right) \tag{E8}$$

Substituting the statistics of Eq. (E6, E7, E8) in Eq. (E5) finally gives Eq. (42) as required.